\documentclass[11pt,a4paper]{article}
\RequirePackage{amsmath}%
\RequirePackage{amsthm}%
\usepackage{amsfonts}%
\usepackage{amssymb}%
\usepackage{graphicx}%

\usepackage{url}
\newtheorem{example}{Example}

\renewcommand{\Re}{\mathop{\mathrm{Re}}}
\renewcommand{\Im}{\mathop{\mathrm{Im}}}

\renewcommand{\i}{\mathrm{i}}

\renewcommand{\Re}{\mathop{\mathrm{Re}}}
\renewcommand{\Im}{\mathop{\mathrm{Im}}}
\renewcommand{\i}{\mathrm{i}}

\newcommand{\bM}{{\bf M}}
\newcommand{\bN}{{\bf N}}
\newcommand{\bI}{{\bf I}}

\begin{document}

\title{External Potential Flow Around Multiple Aerofoils}

\author{Mohamed M.S. Nasser$^{\rm a}$ and Munira Ismail$^{\rm b}$}

\date{}
\maketitle

\vskip-0.8cm %
\centerline{$^{\rm a}$Department of Mathematics, Faculty of Science, King Khalid University,} %
\centerline{P. O. Box 9004, Abha 61413, Saudi Arabia.}%
\centerline{E-mail: mms\_nasser@hotmail.com}
\centerline{$^{\rm b}$Department of Mathematical Sciences, Faculty of Science,} %
\centerline{Universiti Teknologi Malaysia, 81310 UTM Johor Bahru, Johor, Malaysia}%
\centerline{E-mail: mi@mel.fs.utm.my}

\begin{center}
\begin{quotation}
{\noindent {\bf Abstract.\;\;}%
This chapter present a fats boundary integral equation method for numerical 
computing of uniform potential flow past 
multiple aerofoils. The presented fast multipole-based iterative solution 
procedure requires only $O(nm\ln n)$ 
operations where $m$ is the number of aerofoils and $n$ is the number of 
nodes in the discretization of each 
aerofoil's boundary. We demonstrate the performance of our methods on 
several numerical examples. 
}
\end{quotation}
\end{center}
\begin{center}
\begin{quotation}
\end{quotation}
\end{center}


\section{Introduction}
\label{sc:int}

In this chapter,
we consider the two dimensional, steady-state, irrotational flow around multiple 
aerofoils of general shape. We assume that the fluid is incompressible and 
free from viscosity and the boundaries of the aerofoils are stationary and 
impervious. The problem will be solved using
a fast boundary integral method based on the boundary integral equation
with the generalized Neumann kernel presented in~\cite{Nas-cmft11}. 
The integral equation will be solved using the fast method 
presented in~\cite{Nas-fast} which is based on discretizing the integral
equation using the Nystr\"om method with 
the trapezoidal rule then solving the obtained linear system by the generalized 
minimal residual (GMRES) method~\cite{Saa-Sch}. The GMRES method will 
converge significantly faster since the eigenvalues of the coefficient matrix of the 
linear system are clustered around~$1$ (see~\cite{Nas-siam13,Nas-Mur12,Nas-amc11}). 
Each iteration of the GMRES method requires a matrix-vector product which can be 
computed using the Fast Multipole Method (FMM) in $O(mn)$ operations where $m$ 
is the number of aerofoils and $n$ is the number of nodes in the discretization 
of each aerofoil's boundary. Computing
the right-hand side of the integral equation requires applying the \verb|FFT| 
for each of the $m$ aerofoils which requires $O(mn\ln n)$ operations. 
Thus, the complexity of the presented method is $O(mn\ln n)$.

Three numerical examples will be presented. The numerical results illustrate 
that the present method has the ability to handle regions with very high connectivity 
and complex geometry.

\section{Notations and auxiliary material}
\label{sc:aux}

\begin{figure}[ht]%
\centerline{\scalebox{0.7}[0.35]{\includegraphics{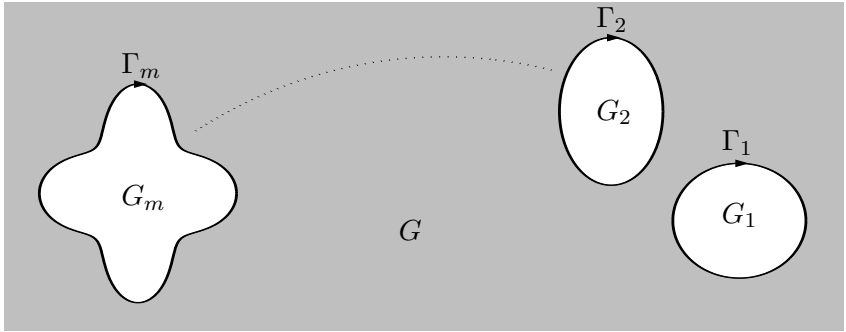}}}
 \vskip-4.50cm \noindent\hspace{9.60cm} $\Gamma_2$
 \vskip+0.75cm \noindent\hspace{9.60cm} $G_2$
 \vskip-1.10cm \noindent\hspace{3.35cm} $\Gamma_m$
 \vskip+0.55cm \noindent\hspace{11.25cm} $\Gamma_1$
 \vskip+0.25cm \noindent\hspace{3.35cm} $G_m$
 \vskip-0.25cm \noindent\hspace{11.25cm} $G_1$
 \vskip-0.25cm \noindent\hspace{7.00cm} $G$
 \vskip+1.00cm
\caption{\rm An unbounded multiply connected region $G$ of connectivity $m$.} 
\label{f:unbd}
\end{figure}

We consider an unbounded multiply connected regions $G$ in the extended complex 
plane $\overline{\mathbb{C}}$ exterior to $m\ge1$ simply connected regions $G_j$, 
$j=1,2,\ldots,m$. We assume that the region $G$ is filled with an irrotational 
incompressible fluid flow and the bounded regions $G_j$, $j=1,2,\dots,m$, represents 
$m$ aerofoils or obstacles in the flow path. 
We assume that the boundaries $\Gamma_j:=\partial G_j$ of the aerofoils are smooth closed Jordan curves. The orientation of the whole boundary $\Gamma:=\partial G=\cup_{j=1}^{m}\Gamma_j$ is such that $G$ is always on the left of $\Gamma$, i.e., the curves $\Gamma_1,\dots,\Gamma_m$ always have clockwise orientations (see Fig.~\ref{f:unbd}). 

The curve $\Gamma_j$ is parametrized by a $2\pi$-periodic twice continuously differentiable complex function $\eta_j(t)$ with non-vanishing first derivative
\begin{equation}\label{e:par}
\dot\eta_j(t)=d\eta_j(t)/dt\ne 0, \quad t\in J_j:=[0,2\pi], \quad j=1,2,\ldots,m.
\end{equation}
We define the total parameter domain $J$ as the disjoint union of the intervals $J_j$. Hence, a parametrization of the
whole boundary $\Gamma$ is defined as the complex function $\eta$ defined on
$J$ by
\begin{equation}\label{e:eta}
\eta(t):= \left\{
\begin{array}{l@{\hspace{1cm}}l}
\eta_1(t), &  t\in J_1,     \\
\vdots     &              \\
\eta_m(t), &  t\in J_m.     \\
\end{array}\right.
\end{equation}

The definition of the function $\eta(t)$ in~(\ref{e:eta}) means that for a given real number $\hat t\in[0,2\pi]$, to evaluate $\eta(\hat t)$, we should know in advance the interval $J_j$ to which $\hat t$ belongs,
i.e., we should know the boundary $\Gamma_j$ contains the point $\eta(\hat t)$, 
then we compute $\eta(\hat t)=\eta_j(\hat t)$.

The real kernel $N$ defined by
\begin{equation}\label{e:N}
 N(s,t) =  \frac{1}{\pi}\Im\left(
 \frac{\dot\eta(t)}{\eta(t)-\eta(s)}\right).
\end{equation}
is known as the Neumann kernel. It is special case of the 
generalized Neumann kernel with $A=1$. 
The kernel $N$ is continuous with
\begin{equation}\label{e:N-d}
N(t,t)= \frac{1}{2\pi}\Im\frac{\ddot\eta(t)}{\dot \eta(t)}.
\end{equation}
The real kernel $M$ defined by
\begin{equation}
 \label{e:M}
 M(s,t) =  \frac{1}{\pi}\Re\left(
 \frac{\dot\eta(t)}{\eta(t)-\eta(s)}\right),
\end{equation}
has a cotangent singularity type. When  $s,t\in J_j$  are 
in the same parameter interval $J_j$, then
\begin{equation}\label{e:M-Mh}
M(s,t)= -\frac{1}{2\pi} \cot \frac{s-t}{2} +M_1(s,t)
\end{equation}
with a continuous kernel $M_1$ which takes on the diagonal the values
\begin{equation}\label{e:M-Mh-tt}
M_1(t,t)= \frac{1}{2\pi}\Re \frac{\ddot\eta(t)}{\dot \eta(t)}.
\end{equation}
See~\cite{Hen3,Mur-Bul,Nas-ibb,Nas-amc11,Weg-Mur-Nas,Weg-Nas} for more details.

We define the Fredholm integral operator with the kernel $N$ and the 
singular operator with the kernel $M$ by
\begin{eqnarray}
 \label{e:bN}
 \bN \mu &:=& \int_J N(s,t) \mu(t) dt, \\
 \label{e:bM}
 \bM \mu &:=& \int_J M(s,t) \mu(t) dt.
\end{eqnarray}
The integral in~(\ref{e:bM}) is a principal value integral.

\section{The external potential flow problem}
\label{sc:flow}

Suppose that $F(z)$ is the complex potential and $W(z)=F'(z)$ is the complex velocity of the flow where $z=x+\i y\in G\cup\Gamma$. The associated velocity field $(u,v)$ is given, in complex form, by the relation
\[
u(x,y)-\i v(x,y) = W(z).
\]
The velocity potential $\phi(x,y)$ and the stream function $\psi(x,y)$ associated with the flow are defined by
\[
\phi(x,y)+\i\psi(x,y)=F(z).
\]
The families of curves
\[
\phi(x,y)=\mathrm{constant}, \quad \psi(x,y)=\mathrm{constant}
\]
are known as the equi-potential curves and the stream lines, respectively,~\cite[p.~98]{Kam07}.  

The complex potential $F(z)$ can be written in the from
\begin{equation}\label{e:Fz}
F(z)=e^{-\i\alpha}z-\i f(z)+\sum_{j=1}^m \frac{\chi_j}{2\pi\i}\log(z-a_j)+c, \quad z\in G\cup\Gamma,
\end{equation}
where $f(z)$ is an analytic function in $G$ with $f(\infty)=0$, 
$c$ is a complex constant and $\chi_j$ is the circulation of the fluid along the boundary component $\Gamma_j$
(Note that the boundaries $\Gamma_j$ are assumed to be clockwise oriented). The complex velocity $W(z)$ is given by
\begin{equation}\label{e:W}
W(z)=e^{-\i\alpha}-\i f'(z)+\sum_{j=1}^m \frac{\chi_j}{2\pi\i}\frac{1}{z-a_j}, \quad z\in G\cup\Gamma.
\end{equation}

It is clear from~(\ref{e:W}) that knowing the function $f'(z)$ is sufficient 
to know the velocity function $W(z)$. For the potential function $F(z)$, the 
constant $c$ in~(\ref{e:Fz}) has no effect on the velocity field. Hence, 
to determine the potential function $F(z)$, it is only required to determine 
the auxiliary function $f(z)$. Then, the stream function is given by
\begin{equation}\label{e:str}
\psi(x,y)=\Im[e^{-\i\alpha}z]-\Re[f(z)]
-\sum_{j=1}^m \frac{\chi_j}{2\pi}\ln|z-z_j|+{\rm constant}
\end{equation}
and the velocity potential is given by
\begin{equation}\label{e:pot}
\phi(x,y)=\Re[e^{-\i\alpha}z]+\Im[f(z)]
+\sum_{j=1}^m \frac{\chi_j}{2\pi}\arg(z-z_j)+{\rm constant}.
\end{equation}

\section{The integral equation with the Neumann kernel}

The boundary values of the analytic function $f(z)$ in~(\ref{e:Fz}) are given by~\cite{Nas-cmft11}
\begin{equation}\label{e:f=g+h+im}
f(\eta(t))=\gamma(t)+h(t)+\i\mu(t)
\end{equation}
where the function $\gamma(t)$ is defined on $J$ by
\begin{equation}\label{e:gam=}
\gamma =\Im[e^{-\i\alpha}\eta]-\sum_{j=1}^m \frac{\chi_j}{2\pi}\ln|\eta-z_j|,
\end{equation}
the function $\mu$ is the unique solution of the integral equation
\begin{equation}\label{e:ie}
\mu-\bN\mu=-\bM\gamma,
\end{equation}
and the function $h$ is given by
\begin{equation}\label{e:h}
h=[\bM\mu-(\bI-\bN)\gamma]/2.
\end{equation} 
The function $h$ is a piecewise constant real-valued function, i.e.,
\begin{equation}\label{e:h-0-m}
h(t)= \left\{
\begin{array}{l@{\hspace{1cm}}l}
h_{1}, &  t\in J_{1},  \\
\vdots     &              \\
h_m, &  t\in J_m,     \\
\end{array}\right.
\end{equation}
with real constants $h_1$, \ldots, $h_m$.

In view of~(\ref{e:f=g+h+im}), the values of the function $f(z)$ can be calculated for $z\in G$ by the Cauchy integral formula
\begin{equation}\label{e:f-cau}
f(z)=\frac{1}{2\pi\i}\int_\Gamma\frac{\gamma+h+\i\mu}{\eta-z}d\eta.
\end{equation}
Then, the stream function can be computed from~(\ref{e:str}).

\section{Numerical examples}

Since the functions $A_j$ and $\eta_j$ are $2\pi$-periodic, 
a reliable procedure for solving the integral equation~(\ref{e:ie}) 
numerically is by using the Nystr\"om method with the trapezoidal 
rule~\cite{Atk97}. Thus solving the integral equation reduces to 
solving an $mn\times mn$ linear system where $m$ is the multiplicity 
of the multiply connected region and $n$ is the number of nodes in 
the discretization of each boundary component.  
Since the integral equation~(\ref{e:ie}) is uniquely solvable, then for 
sufficiently large $n$, the obtained linear system is also 
uniquely solvable~\cite{Atk97}. 
See~\cite{Nas-cmft09,Nas-siam09,Nas-cmft11,Nas-jmaa11,Nas-jmaa13}
for more details.  

In this chapter, the MATLAB function \verb|FBIE| 
presented in~\cite{Nas-fast} will be used to solve the integral 
equation~(\ref{e:ie}) and compute the function $h$ in~(\ref{e:h}). 
The MATLAB function \verb|FBIE| is based on discretizing the integral
equation~(\ref{e:ie}) using the Nystr\"om method with the trapezoidal 
rule then solving the obtained 
linear system by the MATLAB function \verb|gmres|. 
The function \verb|gmres| can be used with a matrix-vector product 
function, i.e., it is not necessary to have an explicit form of the coefficient 
matrix of the linear system. In~\cite{Nas-fast},
the matrix-vector product function for the coefficient matrix of our linear 
system was defined using the function \verb|zfmm2dpart| in the MATLAB toolbox 
\verb|FMMLIB2D| developed by Greengard and Gimbutas~\cite{Gre-Gim12}. 
Thus, the obtained linear systems will be solved in $O(mn)$ operation. 
However, computing
the right-hand side of the integral equation requires applying the \verb|FFT| 
for each of the $m$ boundary components which requires $O(mn\ln n)$ 
operations. 
Thus, the complexity of the presented method is $O(mn\ln n)$.

By solving the integral equation~(\ref{e:ie}) numerically, we obtain an approximation to the boundary values of the function $f$ by~(\ref{e:f=g+h+im}). Then an approximation $f_n(z)$ to the values of the function $f(z)$ for points $z\in G$ will be computed using the Cauchy integral formula~(\ref{e:f-cau}). The integral in~(\ref{e:f-cau}) is discretized by the trapezoidal rule. The FMM will be used for fast computing of the values of $f_n(z)$. See~\cite{Nas-fast} for more details.

\begin{figure}[ht]%
\centerline{
\scalebox{0.5}[0.5]{\includegraphics{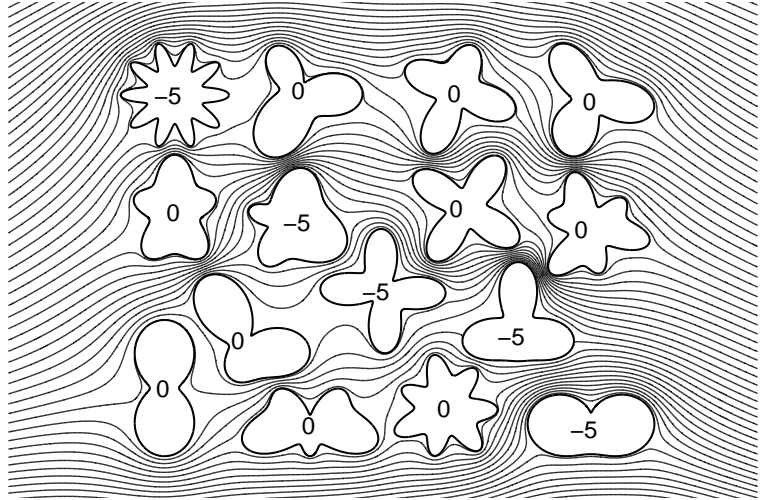}}
}
\caption{\rm Numerical results for Example~\ref{ex:1} obtained with $n=2048$ 
(total number of nodes is $30720$). The streamlines are shown for $\alpha=0$ and for zero circulations along $10$
aerofoils and non-zero circulations along $5$ aerofoils (the circulation along each aerofoils 
is shown inside the curve).} 
\label{f:ex1}
\end{figure}

\begin{example}\label{ex:1}
 The region $G$ is an unbounded multiply connected region exterior to $15$
smooth Jordan curves (see Fig.~\ref{f:ex1}).
\end{example}

\begin{figure}[ht]%
\centerline{
\scalebox{0.5}[0.5]{\includegraphics{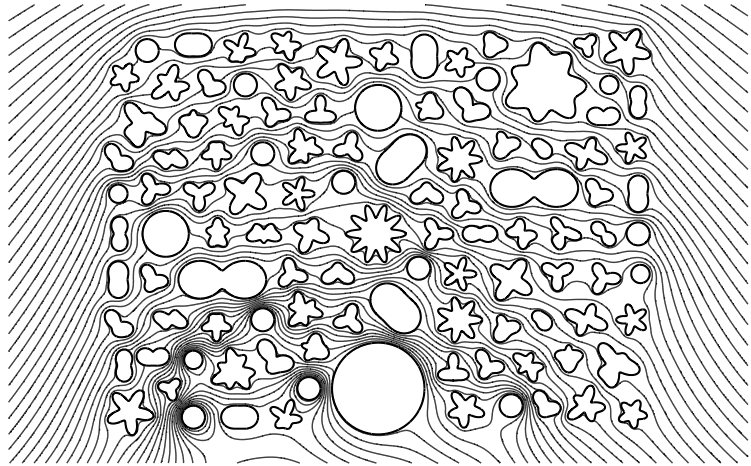}}
}
\caption{\rm Numerical results for Example~\ref{ex:2} obtained with $n=2048$ 
(total number of nodes is $225280$). 
The streamlines  are shown for $\alpha=0$ and for circulations $-5$ along the
cylindrical aerofoils and zero circulations along the other aerofoils.} 
\label{f:ex2}
\end{figure}

\begin{example}\label{ex:2}
The region $G$ is an unbounded multiply connected region exterior to $110$
smooth Jordan curves (see Fig.~\ref{f:ex2}).
\end{example}

\begin{figure}[ht]%
\centerline{
\scalebox{0.5}[0.5]{\includegraphics{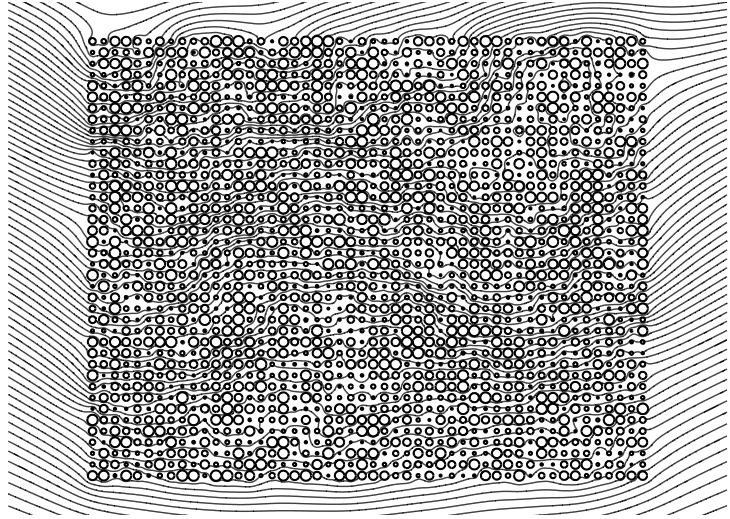}}
}
\caption{\rm Numerical results for Example~\ref{ex:3} obtained with $n=1024$ 
(total number of nodes is $2048000$). 
The streamlines  are shown  for $\alpha=0$ and the circulations along each aerofoil 
is an arbitrary number between $-1$ and $1$.} 
\label{f:ex3}
\end{figure}

\begin{example}\label{ex:3}
The region $G$ is an unbounded multiply connected region exterior to $2000$
circles (see Fig.~\ref{f:ex3}).
\end{example}

\begin{thebibliography}{10}

\bibitem{Atk97}
K.~E. Atkinson.
\newblock {\em The Numerical Solution of Integral Equations of the Second
  Kind}.
\newblock Cambridge University Press, Cambridge, 1997.

\bibitem{Gre-Gim12}
L.~Greengard and Z.~Gimbutas.
\newblock {\em FMMLIB2D: A MATLAB toolbox for fast multipole method in two
  dimensions}.
\newblock Version 1.2, 2012.
\newblock \url{http://www.cims.nyu.edu/cmcl/fmm2dlib/fmm2dlib.html}.

\bibitem{Hen3}
P.~Henrici.
\newblock {\em Applied and Computational Complex Analysis, Vol. 3}.
\newblock John Wiley, New York, 1986.

\bibitem{Kam07}
T.~Kambe.
\newblock {\em Elementary Fluid Mechanics}.
\newblock World Scientific, Singapore, 2007.

\bibitem{Mur-Bul}
A.~H.~M. Murid and M.~M.~S. Nasser.
\newblock Eigenproblem of the generalized neumann kernel.
\newblock {\em Bulletin of the Malaysian Mathematical Science Society},
  26:13--33, 2003.

\bibitem{Nas-cmft09}
M.~M.~S. Nasser.
\newblock A boundary integral equation for conformal mapping of bounded
  multiply connected regions.
\newblock {\em Comput. Methods Funct. Theory}, 9:127--143, 2009.

\bibitem{Nas-siam09}
M.~M.~S. Nasser.
\newblock Numerical conformal mapping via a boundary integral equation with the
  generalized neumann kernel.
\newblock {\em SIAM J. Sci. Comput.}, 31(3):1695--1715, 2009.

\bibitem{Nas-ibb}
M.~M.~S. Nasser.
\newblock The riemann-hilbert problem and the generalized neumann kernel on
  unbounded multiply connected regions.
\newblock {\em The University Researcher (IBB University Journal)}, 20:47--60,
  2009.

\bibitem{Nas-cmft11}
M.~M.~S. Nasser.
\newblock Boundary integral equations for potential flow past multiple
  aerofoils.
\newblock {\em Comput. Methods Funct. Theory}, 11:375--394, 2011.

\bibitem{Nas-jmaa11}
M.~M.~S. Nasser.
\newblock Numerical conformal mapping of multiply connected regions onto the
  second, third and fourth categories of koebe's canonical slit domains.
\newblock {\em J. Math. Anal. Appl.}, 382:47--56, 2011.

\bibitem{Nas-jmaa13}
M.~M.~S. Nasser.
\newblock Numerical conformal mapping of multiply connected regions onto the
  fifth category of koebe's canonical slit regions.
\newblock {\em J. Math. Anal. Appl.}, 398:729--743, 2013.

\bibitem{Nas-siam13}
M.~M.~S. Nasser and F.~A.~A. Al-Shihri.
\newblock A fast boundary integral equation method for conformal mapping of
  multiply connected regions.
\newblock {\em SIAM J. Sci. Comput.}, 35(3):A1736--A1760, 2013.

\bibitem{Nas-Mur12}
M.~M.~S. Nasser and A.~H.~M. Murid.
\newblock Numerical experiments on eigenvalues of the generalized neumann
  kernel.
\newblock In A.H.M. Murid and Y.~Yaacob, editors, {\em Advances in Group
  Theory, DNA Splicing and Complex Analysis}, pages 135--158. Penerbit UTM
  Press, 2012.

\bibitem{Nas-amc11}
M.~M.~S. Nasser, A.~H.~M. Murid, M.~Ismail, and E.~M.~A. Alejaily.
\newblock A boundary integral equation with the generalized neumann kernel for
  laplace's equation in multiply connected regions.
\newblock {\em Appl. Math. Comput.}, 217:4710--4727, 2011.

\bibitem{Nas-fast}
M.M.S. Nasser.
\newblock Fast solution of boundary integral equations with the generalized
  neumann kernel.
\newblock In A.H.M. Murid and Y.~Yaacob, editors, {\em Recent Advances on
  Integral Equations with the Generalized Neumann Kernel}. Penerbit UTM Press,
  Submitted.

\bibitem{Saa-Sch}
Y.~Saad and M.~H. Schultz.
\newblock Gmres: A generalized minimum residual algorithm for solving
  nonsymmetric linear systems.
\newblock {\em SIAM J. Sci. Stat. Comput.}, 7(3):856--869, 1986.

\bibitem{Weg-Mur-Nas}
R.~Wegmann, A.~H.~M Murid, and M.~M.~S. Nasser.
\newblock The riemann-hilbert problem and the generalized neumann kernel.
\newblock {\em J. Comput. Appl. Math.}, 182:388--415, 2005.

\bibitem{Weg-Nas}
R.~Wegmann and M.~M.~S. Nasser.
\newblock The riemann-hilbert problem and the generalized neumann kernel on
  multiply connected regions.
\newblock {\em J. Comput. Appl. Math.}, 214:36--57, 2008.

\end{thebibliography}
\end{document}